# Tunable superluminal and subluminal reflected group delay in an air-Weyl semimetal film-Weyl semimetal substrate layered system

Shenping Wang[1], Huayue Zhang[1], Jie Jiang[2], Leyong Jiang[1, *], and Jipeng Wu[3, †]

*Abstract*—In this work, we theoretically investigate the superluminal and subluminal reflected group delay from an air-Weyl semimetal film-Weyl semimetal substrate layered structure by using 4×4 magneto-optical matrix. We show a tunable transition between positive and negative group delays of reflection pulse in such a layered structure controlled by the properties of the Weyl Semimetal layer and reveal its mechanism to control the propagation properties of the light pulse reflected from such structure. It is demonstrated that the reflected group delays are tunable from positive to negative values and vice versa, which are adjusted by tuning the incident angle, thickness of Weyl semimetal film, tilt parameter, and Fermi energy. The control of the magnitude and the sign of the reflected group delay of light propagation represent a key point in slow and fast light technologies. Our results are helpful to control the pulse propagations and are useful for design of Weyl semimetal-based delay devices.

*Keywords*—Group Delay, Weyl semimetal, Magneto-optical matrix.

## I. Introduction

Pulse group delay represents the speed of phase change with frequency, which is a typical optical phenomenon in the process of beam propagation in medium. It is generally believed that when the group delay is lower than the speed of light in a vacuum, it is called slow light, on the contrary, it is called fast light [1]. The control of group delay of optical pulses has been widely studied in theory. It is generally believed that group delays have extensive and potential applications in many fields, such as efficient storage [2], delay line [3] and all-optical buffers [4]. In this context, researchers are committed to finding ways to achieve enhanced group delay based on a variety of novel materials and combined with various mechanisms and structures. For example, transmission or reflection group delays in structures such as the electromagnetically induced transparency (EIT) system [5], weakly absorbing dielectric [6], Fabry Perot cavity structure [7] and metamaterial [8] have attracted attention. In the experiment, Longhi *et al*. First observed reflected optical pulses from a double Lorentzian fiber Bragg grating [9], while Yao *et al*. Realized a strong broadband negative group delay in a circular waveguide with asymmetric cross-shaped slotted structure in the millimeter-wave region[10]. However, due to the limitations of materials and mechanisms, the optical pulse group delay generated by traditional materials and structures is relatively small, and the regulation characteristics are not sufficiently flexible. With the rapid development of nano photonics, the enhancement and regulation of optical delay in new materials and structures has become the main research direction in the field of group delay.

In recent years, the topological quantum state of matter has become an important and developing field in condensed matter physics and material physics [11-12]. in modern times, topological Semimetals include Weyl Semimetals [20], Dirac Semimetals [21], triple emphasis Semimetals [22], and nodal line Semimetals [23]. Among them, Weyl semimetal (WSM) is a kind of topological material that breaks the temporal inversion symmetry or spatial inversion symmetry, which is protected by the topology and symmetry of the system [13-15]. The landmark feature of WSM is the Fermi arc and the Weyl point in the energy band structure connected to it [16], such as TaAs [17], TaP[18], NbAs [19] have been proven to be WSMs by experiments. WSM is a Dirac-like low-energy dispersion material with analogous Hamiltonian, and the Dira-like energy dispersion can give rise to excellent optoelectronic properties, such as tunable terahertz plasma resonance [24] and the exceptional optical nonlinearity [25-27]; Weyl Semimetals also have unconventional quantum Hall effect [13], epsilon-near-zero response [28], exceptionally high electrical conductivity [29] and other excellent properties, so it has rapidly become one of the research hotspots of condensed matter physics in recent years [30]. At the same time, the unique three-dimensional gapless linear dispersion band structure of Weyl semimetal makes it have many unique properties, such as chiral anomaly, chiral magnetic effect, anti-weak localization, chiral Landau

Manuscript received December 6, 2021. (Write the date on which you submitted your paper for review.) This work was supported by the National Natural Science Foundation of China (Grant No. 11704119), the Hunan Provincial Natural Science Foundation of China (Grant No. 2018JJ3325), Scientific Research Fund of Hunan Provincial Education Department (Grant No. 21B0048) and National College Students' innovation and entrepreneurship training program (Grant No. 202110542014)

[1]School of Physics and Electronics, Hunan Normal University, Changsha 410081, China;
[2]Hunan Key Laboratory of Super Microstructure and Ultrafast Process, School of Physics and Electronics, Central South University, Changsha 410083, China;
[3]College of Railway Transportation, Hunan University of Technology, Zhuzhou 412007,China
Corresponding Author: * jiangly28@hunnu.edu.cn; jipengwu243@163.com

level and negative magnetoresistance effect. This provides more possibilities for better application of Weyl semimetals. In addition, we can adjust the Weyl cone by changing the frequency, tilt parameter, Fermi energy and lattice spacing of the component of the dielectric constant tensor. These characteristics enable us to achieve a large group delay by constructing the composite structure of Weyl semimetal, which provides us with a control means to control the optical properties of Weyl semimetal [31].

In this paper, we are based on the 4 × 4 transfer matrix method, the superluminal and subluminal reflection group delay phenomena of air-Weyl semimetal film-Weyl semimetal substrate structure are studied theoretically, and the calculation results show that the reflection group delay in the layered structure controlled by Weyl semimetal characteristics can realize the flexible switching between positive and negative group delays. The tunable transition between positive and negative group delays reveals the mechanism by which it controls the propagation characteristics of reflected light pulses. We believe that the above results are helpful to the implementation of effective schemes based on delay and the design of Weyl semimetal delay promotion devices.

## II. THEORETICAL MODEL AND METH

We consider using a layered structure. From the perspective of flexible regulation, the structure is composed of Weyl semimetal film (region 1) and Weyl semimetal substrate (region 2), and reflects different structural parameters. We consume a beam of light shining on the layered structure at an incident angle, as shown in Figure 1.

Whether it is a film or a substrate, the dielectric constant tensor $\bar{\bar{\varepsilon}}$ can adopt the following rotation form:

$$\bar{\bar{\varepsilon}} = \begin{bmatrix} \varepsilon_{xx} & \varepsilon_{xy} & 0 \\ \varepsilon_{yx} & \varepsilon_{yy} & 0 \\ 0 & 0 & \varepsilon_{zz} \end{bmatrix}, \quad (1)$$

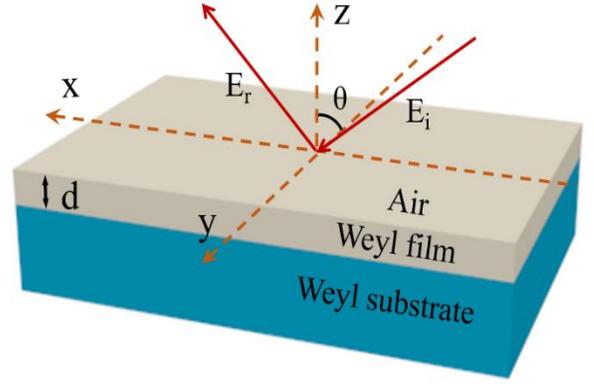

Fig.1. Structural diagram of enhanced group delay realized by Weyl semimetal film and substrate.

here, we mainly consider the dielectric constant component $\varepsilon_{zz}$. For case $|\beta_s| = 0$ where the Weyl cone is not inclined, $\varepsilon_{zz}$ is independent of the tilt parameter $\beta$, the diagonal components are equal to each other, and the component $\varepsilon_{zz}$ can be analytically expressed as [28],

$$\varepsilon_{zz} = 1 + \frac{\alpha}{3\pi}\left[\ln\left|\frac{4\Gamma^2}{4\mu^2 - \omega^2}\right| - \frac{4\mu^2}{\omega^2} + i\pi\Theta(\omega - 2\mu)\right], \quad (2)$$

in this paper, the energy cutoff is set to $\Gamma = 8v_F|Q| \gg (\omega, \mu)$, so we can ignore the influence of $\beta$ on $\Gamma$. Where $\mu$ represents the Fermi energy, $\alpha = e^2/(4\pi\varepsilon_0 \hbar v_F)$, Fermi velocity $v_F = 10^6$ m/s, $\Gamma$ represents the energy cut-off usually related to the tilt parameter $\beta$, and $\Theta(X)$ is the step function. In the case of $|\beta_s| < 1$, $\varepsilon_{zz} = \varepsilon'_{zz} + i\varepsilon''_{zz}$, where $\varepsilon'_{zz}$ and $\varepsilon''_{zz}$ represent its real and imaginary parts:

$$\begin{cases} \varepsilon'_{zz} = 1 + \frac{\alpha\mu^2}{\pi\omega^2}\sum_{s=\pm}\frac{1}{\beta_s^3}\left\{\frac{8}{3}\beta_s - 4\operatorname{arctanh}\beta_s + \ln\left|\frac{4\mu^2 - \omega^2(1+\beta_s)^2}{4\mu^2 - \omega^2(1-\beta_s)^2}\right|\right\} + \frac{\omega^2}{12\mu^2}\sum_{t=\pm}\left[\begin{array}{l}t(1+2t\beta_s)(1-t\beta_s)^2\ln\left|\frac{4\Gamma^2(1-t\beta_s)^2}{4\mu^2 - \omega^2(1-t\beta_s)^2}\right| \\ -\frac{2\mu}{\omega}\left(\frac{4\mu^2}{\omega^2} + 3 - 3\beta_s^2\right)\ln\left|\frac{2\mu - t\omega(1+t\beta_s)}{2\mu + t\omega(1+t\beta_s)}\right|\end{array}\right]\right\}, \\ \varepsilon''_{zz} = \frac{\alpha}{6}\sum_{s=\pm}\Theta\left(\omega - \frac{2\mu}{1+|\beta_s|}\right)\left\{1 - \frac{1}{2}\left[\frac{1 + \frac{3}{2|\beta_s|}\left(\frac{2\mu}{\omega} - 1\right)}{\left(1 - \frac{1}{3\beta_s^2}\left\{\frac{2\mu}{\omega} - 1\right\}\right)^2}\right]\right\}\Theta\left(\frac{2\mu}{1-|\beta_s|} - \omega\right)\right\}. \end{cases} \quad (3)$$

The calculation of the group delay needs to involve the reflection coefficient, so we use 4×4 magneto-optical matrix to calculate the reflection coefficient [31-32]. In order to solve the reflection coefficient of the structure, we introduce the Jones reflection matrix, a 4×4 matrix M [33],

$$M = \left(D^{(0)}\right)^{-1} D^{(1)} P^{(1)} \left(D^{(1)}\right)^{-1} D^{(2)}, \quad (4)$$

the upper index marks air (0), film (1) and substrate (2). In area 0, different modes are independent of each other, so $\left(D^{(0)}\right)^{-1}$ can be expressed as,

$$\left(D^{(0)}\right)^{-1} = \left(2N^{(0)}\cos\theta\right)^{-1} \begin{bmatrix} N^{(0)}\cos\theta & 1 & 0 & 0 \\ N^{(0)}\cos\theta & -1 & 0 & 0 \\ 0 & 0 & N^{(0)} & -\cos\theta \\ 0 & 0 & N^{(0)} & \cos\theta \end{bmatrix}, \quad (5)$$

where $N^{(0)}$ represents the relative refractive index. In area 1, 4×4. The transfer matrix $D^{(1)}$ can be written as,

$$D^{(1)} = \begin{bmatrix} D_{11} & D_{11} & D_{11} & D_{11} \\ D_{21} & -D_{21} & D_{23} & -D_{23} \\ D_{31} & D_{31} & D_{33} & D_{33} \\ D_{41} & -D_{41} & D_{43} & -D_{43} \end{bmatrix}, \quad (6)$$

the corresponding elements in the matrix are
$D_{11} = i\varepsilon_1(\varepsilon_{zz} - N_x^2)$, $D_{21} = i\varepsilon_1 N_{z+}(\varepsilon_{zz} - N_x^2)$,
$D_{23} = i\varepsilon_1 N_{z-}(\varepsilon_{zz} - N_x^2)$, $D_{31} = (\varepsilon_{zz} - N_x^2)(\varepsilon_{xx} - N_x^2 - N_{z+}^2)$,
$D_{33} = (\varepsilon_{zz} - N_x^2)(\varepsilon_{xx} - N_x^2 - N_{z-}^2)$,
$D_{41} = -N_{z+}\varepsilon_{zz}(\varepsilon_{xx} - N_x^2 - N_{z+}^2)$ and
$D_{43} = -N_{z-}\varepsilon_{zz}(\varepsilon_{xx} - N_x^2 - N_{z-}^2)$. Because the film and substrate are Weyl semimetals, and have the same expression. In addition, the propagation in Weyl film can be described by Yeh propagation matrix,

$$P^{(1)} = \begin{bmatrix} e^{i\varphi_+} & 0 & 0 & 0 \\ 0 & e^{-i\varphi_+} & 0 & 0 \\ 0 & 0 & e^{i\varphi_-} & 0 \\ 0 & 0 & 0 & e^{-i\varphi_-} \end{bmatrix}, \quad (7)$$

where $\varphi_\pm = N_{z\pm}d\omega/c$, $N_x = \sin\theta$ and $N_{z\pm} = k_\pm/k_0$, $k_\pm$ can be expressed as:

$$k_\pm^2 = \frac{k_0^2}{2\varepsilon_{zz}}\left[ 2\varepsilon_{xx}\varepsilon_{zz} - (\varepsilon_{xx} + \varepsilon_{zz})N_x^2 \pm \sqrt{(\varepsilon_{zz} - \varepsilon_{xx})^2 N_x^4 - 4\varepsilon_1^2\varepsilon_{zz}N_x^2 + 4\varepsilon_1^2\varepsilon_{zz}^2} \right]. \quad (8)$$

Therefore, we can obtain the expression of each element in matrix M, and the reflection coefficient can be obtained as:

$$\begin{cases} r_{sp} = \dfrac{E_{0s}^r}{E_{0p}^i} = \dfrac{M_{11}M_{23} - M_{21}M_{13}}{M_{11}M_{33} - M_{13}M_{31}}, \\ r_{pp} = \dfrac{E_{0p}^r}{E_{0p}^i} = \dfrac{M_{11}M_{43} - M_{41}M_{13}}{M_{11}M_{33} - M_{13}M_{31}}, \\ r_{ss} = \dfrac{E_{0s}^r}{E_{0s}^i} = \dfrac{M_{21}M_{33} - M_{23}M_{31}}{M_{11}M_{33} - M_{13}M_{31}}, \\ r_{ps} = \dfrac{E_{0p}^r}{E_{0s}^i} = \dfrac{M_{41}M_{33} - M_{43}M_{31}}{M_{11}M_{33} - M_{13}M_{31}}. \end{cases} \quad (9)$$

When TM polarized wave is incident on our structure, the phase of reflection coefficient can also be derived from the expression of $r_{ij}$. We define the group delay as the partial derivative of the phase diagonal frequency, which represents the time required for the wave to complete the reflection process through the structure. Four different delay times can be defined as:

$$\tau_{sp} = \frac{d\phi_{sp}}{d\omega}, \tau_{pp} = \frac{d\phi_{pp}}{d\omega}, \tau_{ss} = \frac{d\phi_{ss}}{d\omega}, \tau_{ps} = \frac{d\phi_{ps}}{d\omega}, \quad (10)$$

where $\phi_{sp}$、$\phi_{pp}$、$\phi_{ss}$ and $\phi_{ps}$ represent the phases of reflection coefficients $r_{sp}$、$r_{pp}$、$r_{ss}$ and $r_{ps}$.

## III. RESULTS AND DISCUSSIONS

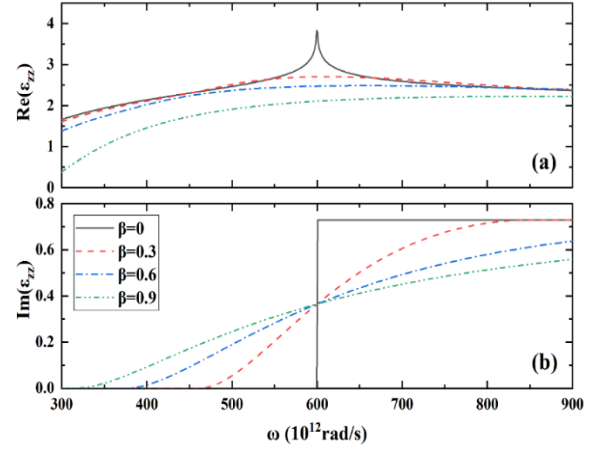

Fig. 2. The component of the permittivity tensor is a function of angular frequency, $\varepsilon_{zz}$ in (a) real part diagram and (b) imaginary part diagram with different tilt parameters. The relevant parameters are $v_F = 10^6$ m/s and $Q = 1$ nm$^{-1}$.

In order to obtain the maximum group delay by calculation, next, in this part, we discuss the delay characteristics of the whole structure. Firstly, we pay attention to the structure composed of air / Weyl semimetal film (region 1) / Weyl semimetal substrate (region 2), and draw the variation diagram of Weyl semimetal dielectric constant component $\varepsilon_{zz}$ with angular frequency, as shown in Fig. 2. In this figure, we plot the function curves of the angular frequency of the real part Fig. 2(a) and imaginary part Fig. 2(b) of the dielectric constant component $\varepsilon_{zz}$ of Weyl semimetal under different tilt parameters. The relevant parameters are $v_F = 10^6$ m/s, $Q = 1$ nm$^{-1}$ and $\Gamma = 8v_F|Q|$. As shown in the figure, when Weyl cone $|\beta| \neq 0$, the dielectric constant component increases monotonically and linearly with the angular frequency. Obviously, the smaller $|\beta|$, the lower the increased amplitude. In particular, for Weyl cone $|\beta| = 0$, such a change trend will change. At $\omega = 600 \times 10^{12}$ rad/s, a transition will occur in the imaginary part of the dielectric constant component, Moreover, the real part of the dielectric constant component will increase first and then decrease. The reason for this phenomenon is that there is a step function in the expression of $\varepsilon_{zz}$ at $|\beta| = 0$. Based on this unique curve, we will reflect more abundant group delay characteristics when calculating group delay. By increasing the inclination angle of the Weyl cone, the dissipative response widens, and the imaginary part value of $\varepsilon_{zz}$ is limited to a larger angular frequency range, which is in the anisotropy with dielectric loss ε- Near zero (ENZ) system is of great significance. Such a characteristic can be used to control beam directivity or electromagnetic wave absorption. In short, the components of the dielectric constant tensor can be tuned by changing, $\mu$ and $\lambda$, so as to realize the feasible tuning of the group delay.

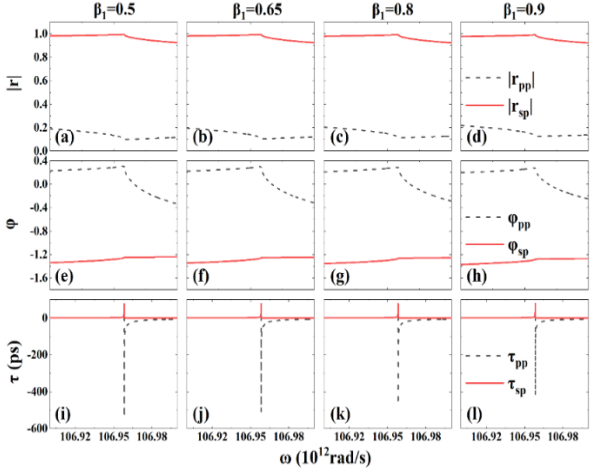

Fig. 3. The group delay with different amplitude and phase of the reflected wave and its corresponding diagram are the tilt parameter $\beta_1 = 0.5$、$0.65$、$0.8$、$0.9$ from left to right. Other parameters are the same as shown in Fig. 2.

After multiple reflection and transmission in the anisotropic layer, the incident wave produces TM and TE reflected and transmitted waves. We mainly study the optical phenomenon of group delay time when the incident light is polarized in TM. In this part, we plot the evolution of reflection, phase and group delay relative to angular frequency before and after different Weyl semimetal film tilt parameters $\beta_1$, as shown in Fig. 3. Considering that the TM polarized wave irradiates the Weyl semimetal film from the air, the mode curve of Fig. 3(a) reflection coefficient has a small change near $\omega = 106.95 \times 10^{12}$ rad/s, so we will study the reflection group delay characteristics of the Weyl semimetal heterostructure near $\omega = 106.95 \times 10^{12}$ rad/s. It is worth noting here that the incident plane wave with TM polarization produces two reflected plane waves and two transmitted plane waves, both of which contain S and P polarized plane waves. Therefore, the solid line and scribed line in the figure are the results related to P and S polarized reflected waves, respectively. Obviously, the reflection of the s-polarized pulse is a positive group delay. With the increase of film tilt parameter $\beta_1$, the reflection group delay parameter gradually increases from 78.2 ps to 79.7 ps, resulting in a weak change; On the contrary, the p polarization pulse reflection is a negative group delay. With the increase of Weyl semimetal film tilt parameter $\beta_1$, it can be clearly found that the p polarization reflection negative group delay gradually increases from $-537.2$ ps to $-419$ ps, which also means that the reflection group delay can be designed by controlling the Weyl semimetal film tilt parameter. Therefore, the tilt parameters of Weyl semimetal films play an important role in determining the group delay. Since the change is obvious under p polarization, we will study the influence of other parameters of Weyl semimetal on group delay under the condition that the incident and reflection are p polarization.

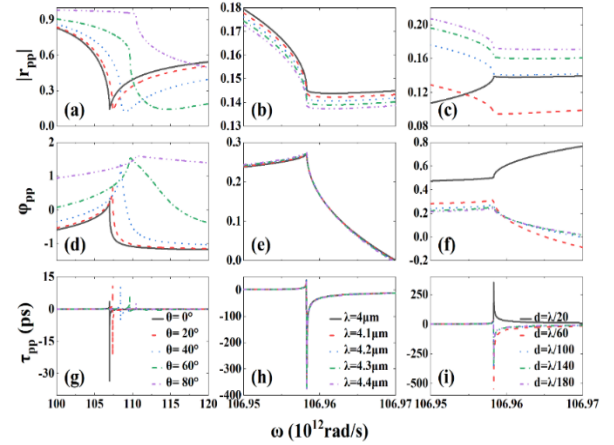

Fig. 4. In different cases, the mode of reflection coefficient and its phase and reflection group delay are typical characteristics. Solid line, dash line, dotted line, dotted line and double dotted line represent $\theta = 0°$、$20°$、$40°$、$60°$、$80°$ on panel (a), (d) and (g), $\lambda = 4\,\mu m$、$4.1\,\mu m$、$4.2\,\mu m$、$4.3\,\mu m$ in (b), (E) and (H) and $d = \lambda/20$、$\lambda/60$、$\lambda/100$、$\lambda/140$、$\lambda/180$ in (c), (f) and (I), respectively.

In this part, Fig.4 shows the properties of reflection, the dependence of its phase and group delay on the diagonal frequency at different incident angles $\theta$ in Fig. 4 (a), 4 (d) and 4 (g), different wavelengths $\lambda$ in Fig. 4 (b), 4 (E) and 4 (H), and different Weyl semimetal film thicknesses $d$ in Fig. 4 (c), 4 (f) and 4 (I). Referring to Figs. 4 (a), 4 (d) and 4 (g), with the change of angle, the change point of the mode of the reflection coefficient gradually moves to the right from $\omega = 106.95 \times 10^{12}$ rad/s to $\omega = 110 \times 10^{12}$ rad/s; The reflection phase slope first increases and then decreases, and the highest point increases gradually from 0.26 to 1.6 with the increase of angle; There are positive and negative group delays in the reflection group delay. With the change of angle, the positive group delay first increases and then decreases, increases from 3.7 ps to 11.2 ps, and then gradually decreases to 2.8 ps. The negative group delay gradually increases from $-33.4$ ps to disappear. In Figs. 4 (b), 4 (E) and 4 (H), with the change of wavelength, the mode of reflection coefficient is always a small change, and with the increase of angular frequency, the minimum value of the mode of reflection coefficient tends to decrease gradually, and the reflection phase is also in a weak change. The corresponding values of reflection positive and negative group delays are 38 ps and $-367$ ps, respectively. Referring to Figs. 4 (c), 4 (f) and 4 (I), with the gradual decrease of Weyl semimetal thickness, at the $d = \lambda/60$ value point, the mode of reflection coefficient first increases to 0.14, then slowly changes to gradually decreases, and then slowly changes; The reflection phase changes slowly from the original to 0.49, then increases to a slow change, and then decreases gradually; Thus, the reflection group delay value changes sharply from $355.14\text{ps}$ to $-563.4\text{ps}$, and then gradually increases to $-307\text{ps}$.

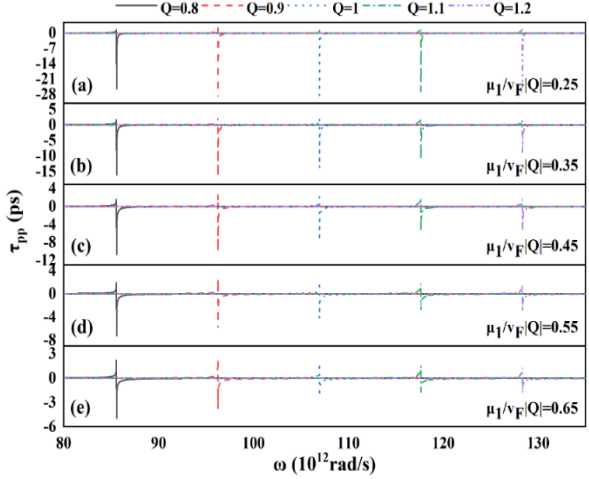

Fig. 5. Comparison of reflection group delays with different lattice spacing at different Fermi levels when the incident angle is $\theta = 0°$ and the thickness of Weyl semimetal film is $d = \lambda/100$.

In the previous discussion, it is known that the reflection group delay is sensitive to the parameters of Weyl semimetal. One of the remarkable characteristics of Weyl semimetals is that they can systematically change their Fermi levels. The Fermi level $\mu$ can move near the charge neutral point by doping, changing the temperature or changing the lattice constant of the material by pressure. In order to study how the changes of lattice spacing $Q$ and Fermi level $\mu$ change the electromagnetic response of Weyl semimetal, the influence of different Fermi levels on the reflection group delay will be discussed. As shown in Fig. 5, the incident angle is $\theta = 0°$, the wavelength is $\lambda = 4.2\,\mu\text{m}$, and the thickness of Weyl semimetal film is $d = \lambda/100$. Three important features can be found in the figure: after determining other parameters, when the lattice spacing is $Q = 0.8\,\text{nm}^{-1}$, the reflection negative group delay increases from $-25.58\text{ps}$ to $-4.9\text{ps}$ with the increase in Fermi level, and the maximum value of group delay corresponding to different $Q$ values moves to the right to the maximum value of $-1.77\text{ps}$. When the Fermi level is fixed, we find that the lattice spacing increases gradually, and the minimum value of the reflection negative group delay shifts to the right. When the Fermi level is $\mu_1/v_F |Q| = 0.25$, the maximum value of the reflection negative group delay is about $-30.01\text{ps}$. Therefore, Weyl semimetal can be easily manipulated by chemical modification or applied voltage, resulting in the change of reflection group delay. Therefore, lattice spacing $Q$ also has an important influence on group delay, and the Fermi level can promote the influence of lattice spacing.

## IV CONCLUSION

In summary, we theoretically study the superluminal and subliminal reflection group delay in the layered structure of air-Weyl semimetal film-Weyl semimetal substrate based on the 4×4 magneto-optical matrix method. The calculation results show that the reflection group delay can be tuned from positive to negative, and vice versa, so as to realize flexible switching between positive and negative. In addition, we also adjust the reflection group delay by adjusting the incident angle, wavelength, Weyl semimetal film thickness, tilt parameters, Fermi energy and lattice spacing, which provides support for the realization of tunable group delay devices. The size and symbol of group delay play an important role in the control of fast and slow light. Therefore, we believe that this scheme can find a feasible application in the field of delay-based devices.

### ACKNOWLEDGMENT

This work was supported by the National Natural Science Foundation of China (Grant No. 11704119), the Hunan Provincial Natural Science Foundation of China (Grant No. 2018JJ3325), Scientific Research Fund of Hunan Provincial Education Department (Grant No. 21B0048) and National College Students' innovation and entrepreneurship training program (Grant No. 202110542014)